\documentstyle[aps,twocolumn]{revtex}
\input epsf
\tightenlines
\begin{document}
\draft
\preprint{VECC/NTH97xxx}
\title{Soft Electromagnetic Radiations From Equilibrating 
Quark-Gluon Plasma}
\author{Dipali Pal}
\vskip 0.4in
\address{ Variable Energy Cyclotron Centre, 1/AF Bidhan Nagar, Calcutta
700 064, India}
\author{Munshi Golam Mustafa{\thanks{Alexander von Humboldt Fellow, and 
on Leave from Saha Institute of Nuclear Physics, 1/AF Bidhannagar, 
Calcutta - 700 064, India}}}
\vskip 0.4in
\address{Institut f\"{u}r Theoretische Physik, Universit\"{a}t Giessen, 
D-35392 Giessen, Germany} 
\date{\today}
\maketitle
\begin{abstract}

\vskip 0.25in

We evaluate the bremsstrahlung production of low mass dileptons and soft
photons from equilibrating and transversely expanding quark gluon plasma 
which may be created in the wake of relativistic heavy ion collisions. 
We use initial conditions obtained from the self screened parton cascade model. 
We consider a boost
invariant longitudinal and cylindrically symmetric transverse expansion 
of the parton plasma and find that for low mass dileptons ( $M \leq 0.3$ 
GeV) and soft photons ( $p_{T} \leq 0.5$ GeV), the bremsstrahlung 
contribution is rather large compared to annihilation process at both
RHIC and LHC energies. We also find an increase by a factor of 15-20
in the low mass dileptons and soft photons yield as one goes from RHIC
to LHC energies.  

\end{abstract}


\vskip 0.25in


\narrowtext
\vskip 0.25in

\section{INTRODUCTION}

The study of ultra-relativistic heavy ion collision explores an
opportunity to verify the possible occurrence of a phase transition from
hadronic matter to deconfined quark matter~ \cite{HM96}. In order to detect
this exotic phenomenon, various signals have been proposed. Among
them, photons and dileptons have long been considered as excellent
probes of the early stages of relativistic heavy ion collision. 
They are the only signals emitted by the Quark Gluon phase that reach
the detector directly without being affected by final state
interaction. As a result, their detection would provide valuable and
reliable information about the moment of their formation. 

One would like to understand several aspects of the matter produced in 
heavy ion collisions, viz.,
how does the initial partonic system evolve and how quickly does it attain
kinetic equilibrium? How quickly, if at all,
does it attain chemical equilibrium?
And finally, how can we uncover the history of this evolution
by studying the spectra of the produced particles, many of which may
decouple from the interacting system only towards the end? These and related
questions have been actively debated in recent times.
Thus, it is believed by now that the large initial parton density  may force
many collisions among the partons in a very short time and lead to a kinetic
equilibrium~\cite{kin}. The question of chemical equilibration~\cite{sspc} 
is more involved, as it depends on the time available to the system. 
If the time available is too short (3--5 fm/c), as at the
energies ($\sqrt{s} \simeq 200$ GeV/nucleon) accessible at the
Relativistic Heavy Ion Collider (RHIC), the QGP will end its journey far
away from chemical equilibrium. At the energies ($\sqrt{s}
\simeq 5.5$ TeV/nucleon) that will be achieved at the CERN Large Hadron
Collider (LHC) this time could be large (more than 10 fm/$c$), driving
the system very close to chemical equilibration~\cite{biro,chem,kg,ja,xmh},
if there is only a longitudinal expansion.
However, this time is also large enough to enable a rarefaction wave from
the surface of the plasma to propagate to the center
($ \tau_s \sim R_T/c_s$; $R_T$ is the transverse nuclear dimension and
$c_s$ is the speed of sound),
 thus introducing large transverse velocity (gradients) in the system.
The large transverse velocities may
impede the chemical equilibration by introducing a faster cooling.
The large velocity gradients may drive the system away from chemical
equilibration by introducing an additional source of depletion of the
number of  partons~\cite{munshi} in a given fluid element.

Dileptons and photons are however produced at every stage of the collision
and in an expanding
system their number is obtained by an integration over the four-volume
of the interaction zone.
At very early times the temperature is rather high and we have a copious
production of large transverse momentum photons, and large mass dileptons. 
This should give us a reliable information about the initial stages 
of the plasma.
However, the transverse flow of the system is moderate at early times,
and their production will only be marginally affected by the flow. By the 
time the flow and the other aspects of QGP develop, the temperature would 
have dropped 
considerably and we have a large production of low transverse momentum
photons and low mass dileptons as well.
However they will have, in addition to the
contributions from the plasma, a large contribution from hadronic
reactions. 
We shall be required to understand them in detail before
we can confidently embark upon the task of deciphering the more involved
development of the QGP,  as it cools (see, e.g.,Ref,~\cite{munshi}).

The soft photons and low mass dileptons, mostly produced at later stages 
when the system is expected to be affected strongly by transverse flow, 
can provide reliable and useful information about the later stage of the 
interaction zone. As a consequence, this soft region has been suggested 
as a possible window for obtaining imprints about the flow. Obviously,
the precise location of this window is determined by the initial 
conditions of the plasma and the dynamics of the space time evolution. 
We may also add that recent data from CERES experiment \cite{ceres} at 
CERN SPS for the $S + Au$ system show a strong excess in the dielectron 
yield compared to $p + Be$ and $p + Au$ systems below the $\rho$ mass. 
This result has already been motivated a lot of systematic studies in 
this mass region \cite{li}. It is generally accepted that bremsstrahlung 
processes mainly contribute at low masses. In Ref. \cite{dip} it has 
been shown that the description of the CERES data will improve to some 
extent, when one adds hadronic bremsstrahlung contribution. 

For a {\it fully equilibrated plasma}, the production of low mass dileptons 
and soft photons due to bremsstrahlung process has been studied in great 
detail in the literature \cite{haglin,dipali,eggers}. However, 
in the recently formulated self-screened parton cascade (SSPC) model~\cite{sspc}
early hard scattering produce a medium which screens the long range 
color fields associated with softer interactions. When two heavy nuclei 
collide at sufficiently high energy, the screening occurs on the length 
scale where perturbative QCD still applies. This approach yields predictions
for the initial conditions of the forming QGP without the need of any
{\it ad-hoc} virtuality and momentum cut-off parameters. These calculation 
also show that the QGP likely to be formed in RHIC and LHC energies could 
be hot
and initially far away from chemical equilibrium. With passage of time, 
chemical reactions among the partons will push it towards chemical 
equilibration initially, but the large transverse velocity gradient 
drives the system away from chemical equilibration~\cite{munshi}. 
It is quite evident 
that signals which are emitted at QGP phase would somehow reflect the 
state of non-equilibration at the time of the phase transition. 

In the present work, we mainly concentrate on a region of low mass 
dileptons ( $M \leq 0.5$ GeV ) and soft photon ( $ p_T \leq 0.5$ GeV ) 
production due to bremsstrahlung in a more complete manner, $i.e.$, in 
a chemically equilibrating and transversely expanding quark-gluon plasma. 
Since the energies of the electromagnetic radiations are low enough,
we obtain the production of low mass dileptons and soft photons from 
bremsstrahlung processes within a soft photon approximation \cite{spa}. 
The goal of this approximation is to separate the process of electromagnetic
radiation from the strong interaction components. One also neglects the
four momentum of photon in the $\delta$-function, which could be 
compensated by inserting a phase-space correction factor (see below). 

It is worth noting here that in a recent calculation Aurenche 
et al.~\cite{auren} have also calculated bremsstrahlung of soft photons
(real and virtual) in the frame work of effective perturbative expansion
based on resummation of hard thermal loops (HTL). They demonstrated that
bremsstrahlung appears only on the two-loop level when the exchange gluon is
space like, and its contribution is of same order as that of one-loop.
Since the exchanged gluon is space-like, it manifests itself in the rate
to the square of the thermal gluon mass ($m_g\sim gT$), in contrast to the 
one-loop result where only the thermal quark mass appears~\cite{baart}. 
For virtual photons this result is of the same order in coupling constant, but
it can differ quantatively because of the gluon mass.
On the other hand, for real photon bremsstrahlung a factor ($1/g^2$) arises
due to collinear singularities associated with fermionic propagator caused 
by the vanishing photon mass in the external line of the vacuum polarization 
diagrams in the two loops. This is compensated by the factor $m^2_g$ ($\sim
g^2 T^2$) 
which leaves the two-loop bremsstrahlung contribution for real soft photon
at the same order in $gT$ as that of one-loop. We shall come back to this
aspect later.

The paper is organized in the following way. In Sec.II, we briefly recall 
hydrodynamic and chemical evolution of the plasma in a transverse direction.
In Sec.III we present the formulation for bremsstrahlung production of soft
photon and low mass dileptons in a chemically equilibrating and transversely 
expanding plasma. We discuss our results in Sec.IV. Finally in Sec.V, we 
give a brief summary.

\section{HYDRODYNAMIC EXPANSION, AND CHEMICAL EQUILIBRATION}

We assume that kinetic equilibrium has been achieved when the momenta of
partons become locally isotropic. At the collider energies it has been 
estimated that, $\tau_i\approx 0.2-0.3$ fm/$c$~\cite{kin}. 
 Beyond this point, further expansion is described by
 hydrodynamic equations and the chemical equilibration is governed by 
a set of master equations which are driven by the two-body reactions
($gg\,\leftrightarrow\,q\bar{q}$) and gluon multiplication and its
inverse process, gluon fusion ($gg\,\leftrightarrow\,ggg$). The other
(elastic) scatterings help maintain thermal equilibrium.
The hot matter continues to expand and cools due to expansion and
chemical equilibration.  We shall somewhat arbitrarily terminate
the evolution once the energy density reaches some critical
value (here taken as $\epsilon_f=1.45$ GeV/fm$^3$ \cite{energy}). 

The expansion of the system is described by the equation for 
conservation of energy and momentum of an ideal fluid:
\begin{equation}
\partial_\mu T^{\mu \nu}=0 \; , \qquad
 T^{\mu \nu}=(\epsilon+P) u^\mu u^\nu + P g^{\mu \nu} \, ,
\label{hydro}
\end{equation}
where $\epsilon$ is the energy density and $P$ is the pressure measured 
in the frame comoving with the fluid. 
The four-velocity vector $u^\mu$ of the 
fluid satisfies the constraint $u^2=-1$. 
For a partially equilibrating plasma the distribution functions for 
gluons and  quarks 
are assumed
\begin{equation}
n_i(E_i,\lambda_i)= \lambda_{i} {\tilde n}_i(E_i) \ \ ,
 \label{befd}
\end{equation}
where ${\tilde n}_i(E_i)=({e^{\beta{E_i}}\mp 1})^{-1}$
is the BE (FD) distributions for gluons
(quarks), and $\lambda_i$ is the fugacity for parton species $i$, which
describes the deviation from chemical equilibrium. This fugacity factor
takes into account undersaturation of parton phase space density,
$i.e.$, $0\leq\lambda_i\leq 1$.
The equation of state for a partially equilibrated plasma of massless 
particles can be written as
\cite{biro}
\begin{equation}
\epsilon=3P=\left[a_2 \lambda_g +  b_2 \left (\lambda_q+\lambda_{\bar q}
\right ) \right] T^4 \, ,
\label{eos}
\end{equation}
where $a_2=8\pi^2/15$, $b_2=7\pi^2 N_f/40$, $N_f $ is 
the number of dynamical quark flavors. Now, the density of an
equilibrating partonic system can be written as 
\begin{equation}
n_g=\lambda_g \tilde{n}_g,\qquad 
 n_q=\lambda_q \tilde{n}_q,
\end{equation}
where $\tilde{n}_k$ is the equilibrium density for the parton species $k$:
\begin{equation}
\tilde{n}_g=\frac{16}{\pi^2}\zeta(3) T^3=a_1 T^3,
\end{equation}
\begin{equation}
\tilde{n}_q=\frac{9}{2\pi^2}\zeta(3) N_f T^3=b_1 T^3.
\end{equation}
We further assume that $\lambda_q=\lambda_{\bar{q}}$.  The equation of
state (\ref{eos}) implies the speed of sound $c_s=1/\sqrt{3}$. We
 solve the hydrodynamic equations (\ref{hydro}) with the assumption
that the system undergoes a boost invariant longitudinal expansion along
the $z$-axis and a cylindrically symmetric transverse expansion \cite{vesa}.
It is then sufficient to solve the problem for $z=0$, because of the
assumption of boost invariance.

The master equations \cite{biro} for the dominant chemical reactions 
$gg \leftrightarrow ggg$ and $gg \leftrightarrow q\bar{q}$  are
\begin{eqnarray}
\partial_\mu (n_g u^\mu)&=&n_g(R_{2 \rightarrow 3} -R_{3 \rightarrow 2})
                    - (n_g R_{g \rightarrow q}
                       -n_q R_{q \rightarrow g} ) \, , \nonumber\\
\partial_\mu (n_q u^\mu)&=&\partial_\mu (n_{\bar{q}} u^\mu)
                     = n_g R_{g \rightarrow q}
                       -n_q R_{q \rightarrow g},
\label{master1}
\end{eqnarray}
in an obvious notation.
In case of transverse expansion, the  master equations  can be shown
\cite{munshi} to lead to partial differential equations:
\begin{eqnarray}
\frac{\gamma}{\lambda_g}\partial_t \lambda_g &+& \frac{\gamma v_r }{\lambda_g}
\partial_r \lambda_g +\frac{1}{T^3}\partial_t (\gamma T^3) + \frac{v_r}{T^3}
\partial_r (\gamma T^3)  \nonumber\\ 
&+& \gamma \partial_r v_r +\gamma \left( \frac{v_r}{r}+\frac{1}{t}\right) 
\nonumber\\ &=& 
R_3 ( 1- \lambda_g ) -2 R_2 \left( 1-\frac{\lambda_q \lambda_{\bar{q}}}
{\lambda_g^2}\right) \, , 
\nonumber\\
\frac{\gamma}{\lambda_q}\partial_t \lambda_q &+& \frac{\gamma v_r }{\lambda_q}
\partial_r \lambda_q +\frac{1}{T^3}\partial_t (\gamma T^3) + \frac{v_r}{T^3}
\partial_r (\gamma T^3) \nonumber\\
&+& \gamma \partial_r v_r +\gamma \left( \frac{v_r}{r}+\frac{1}{t}\right) 
\nonumber\\ &=&
R_2 \frac{a_1}{b_1} \left(
\frac{\lambda_g}{\lambda_q}-\frac{\lambda_{\bar{q}}}{\lambda_g}\right)\, ,
\label{master}
\end{eqnarray}
where $v_r$ is the transverse velocity and $\gamma=1/\sqrt{1-v_r^2}$.
The $R_2$ and $R_3$ related to the rates appearing in (\ref{master1}) are
given by,
\begin{eqnarray}
R_2 & \approx & 0.24 N_f \alpha_s^2 \lambda_g T \ln (1.65/\alpha_s \lambda_g),
\nonumber\\
R_3 & = & 1.2 \alpha_s^2 T (2\lambda_g-\lambda_g^2)^{1/2},
\end{eqnarray}
where the color Debye screening and the Landau - Pomeranchuk - Migdal effect
suppressing the induced gluon radiation have been taken into account,
explicitly.

The hydrodynamic equations (\ref{hydro}) are solved numerically, with 
the initial conditions obtained from the SSPC model
\cite{sspc}, to get $\epsilon(r,t)$ and $v_r(r,t)$, which serve as input 
into the equations (\ref{master}) for the fugacities (for details see
Ref.~\cite{munshi}). For the convenience of reader, we plot the different
parton fugacities in Figs.(1-2) for a transversely expanding plasma likely
to be produced in RHIC and LHC collider energies.  In the figures $N$ 
denotes the time.

\begin{figure}
\epsfxsize=3.25in
\epsfbox{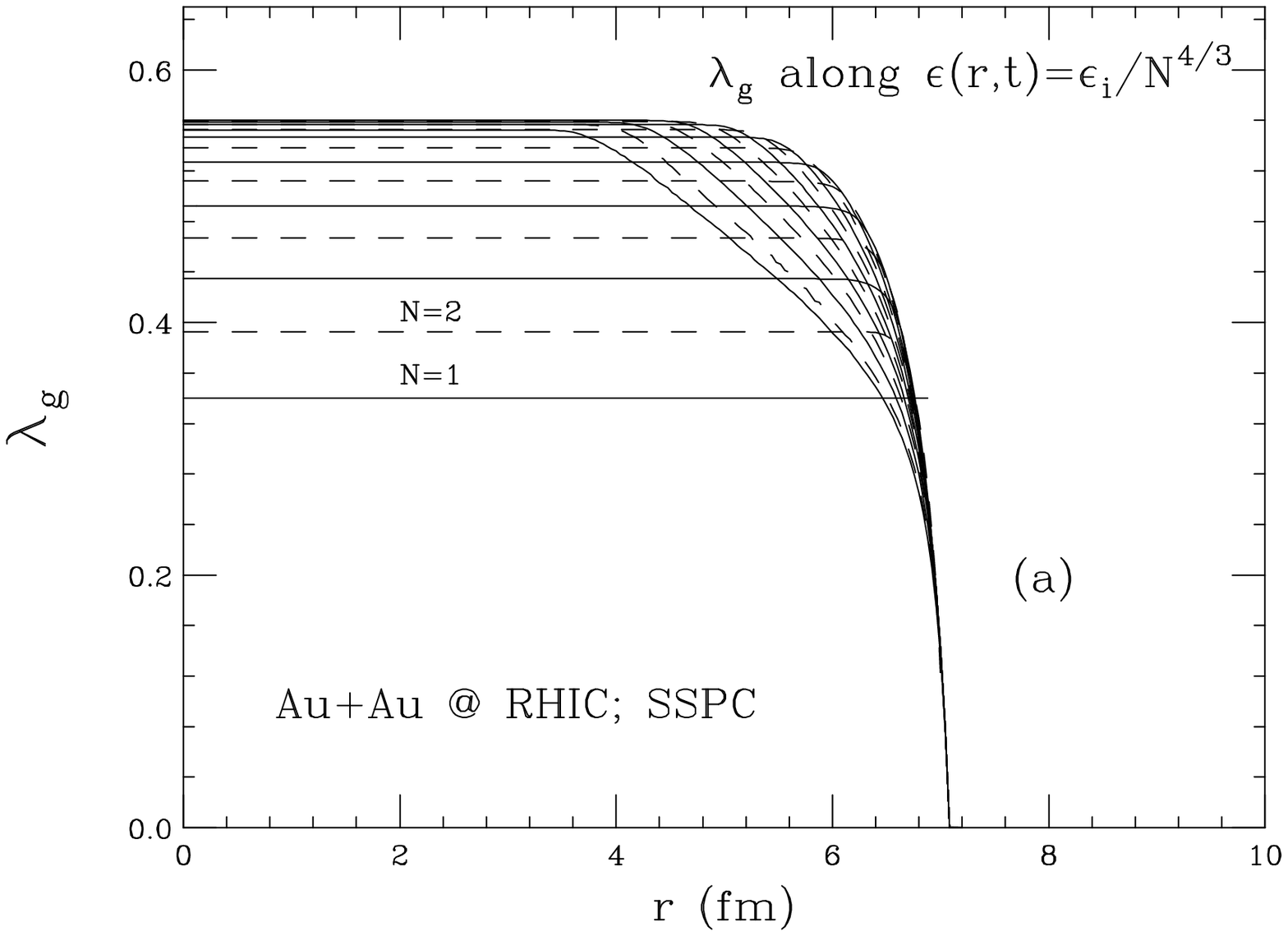}
\vskip 0.10in
\epsfxsize=3.25in
\epsfbox{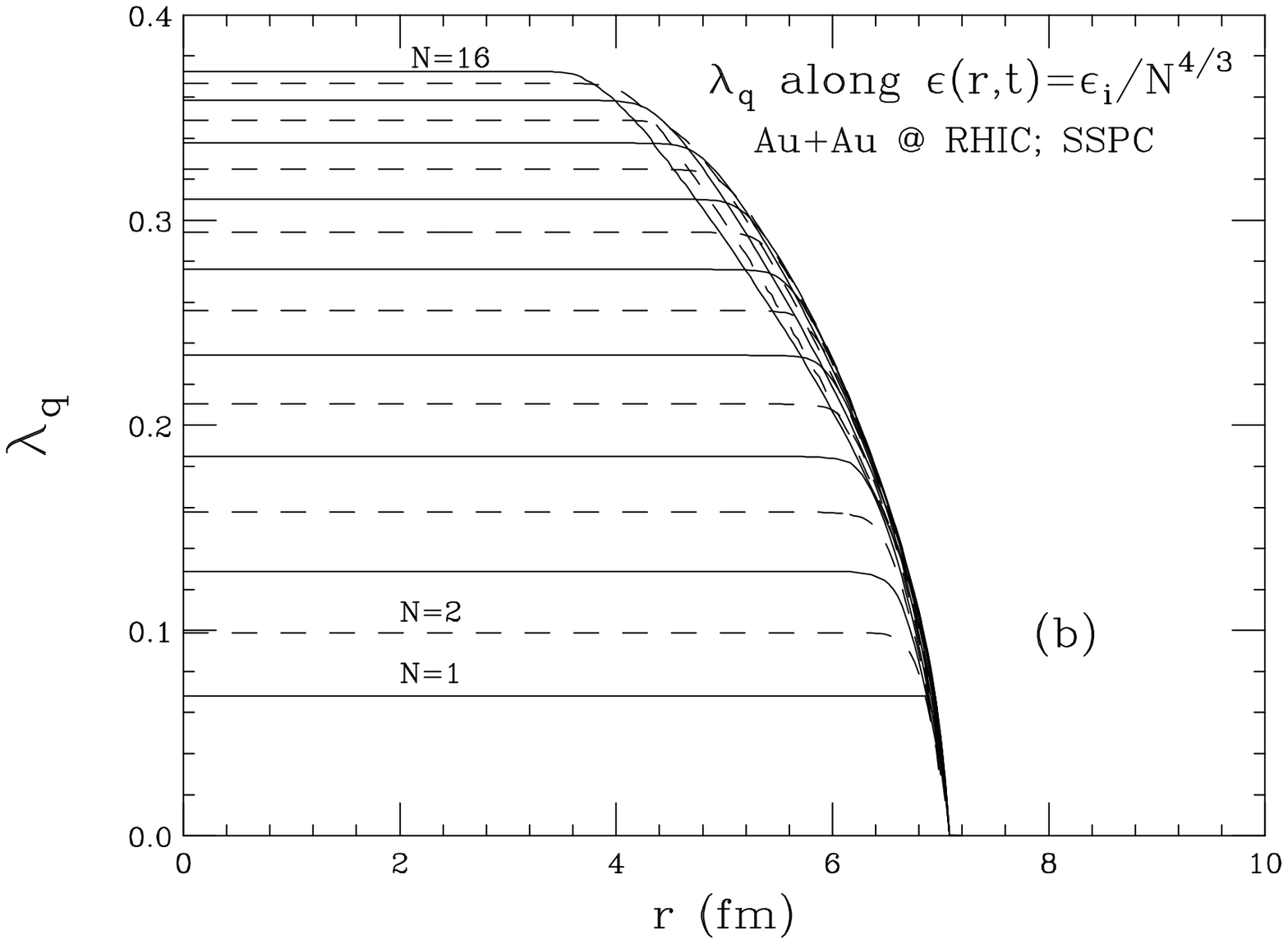}
\vskip 0.10in
\caption{
(a) Gluon fugacities  and (b) Quark fugacities along the
constant energy density contours for RHIC energies. The constant energy
density contours are the solution [9]
of the hydrodynamic 
equation (1)
 with the initial conditions obtained from SSPC 
model.
} 
\end{figure}

One can see that the quark fugacities always lag the gluon 
fugacities (see Figs.1 and 2). This is obvious as SSPC model predicts the 
gluon 
dominated plasma. At both the energies the fugacities initially increase
with time (denoted by $N$), then start decreasing because by then 
the transverse expansion has set in ($v_r$ is non-zero). Thus the region
where the transverse velocity is large (later time), the plasma ends it 
journey farther away from chemical equilibrium. The reason is
the velocity gradient causes a depletion in the partons number in the fluid
element, which could not be made up by the number of partons produced due
to parton chemistry~\cite{munshi}, resulting in a decrease of the parton 
fugacities. The chemical equilibration cannot be achieved at RHIC as the life
time (3-4 fm/$c$) is too small. Due to the longer life time of the plasma
phase at LHC energies (12 fm/$c$), the effects of the transverse expansion
are more dramatic, and the entire matter participate in the flow.
There is a possibility of approaching chemical equilibration at LHC energies,
and goes far away from it as velocity gradient is substantial. 

\begin{figure}
\epsfxsize=3.25in
\epsfbox{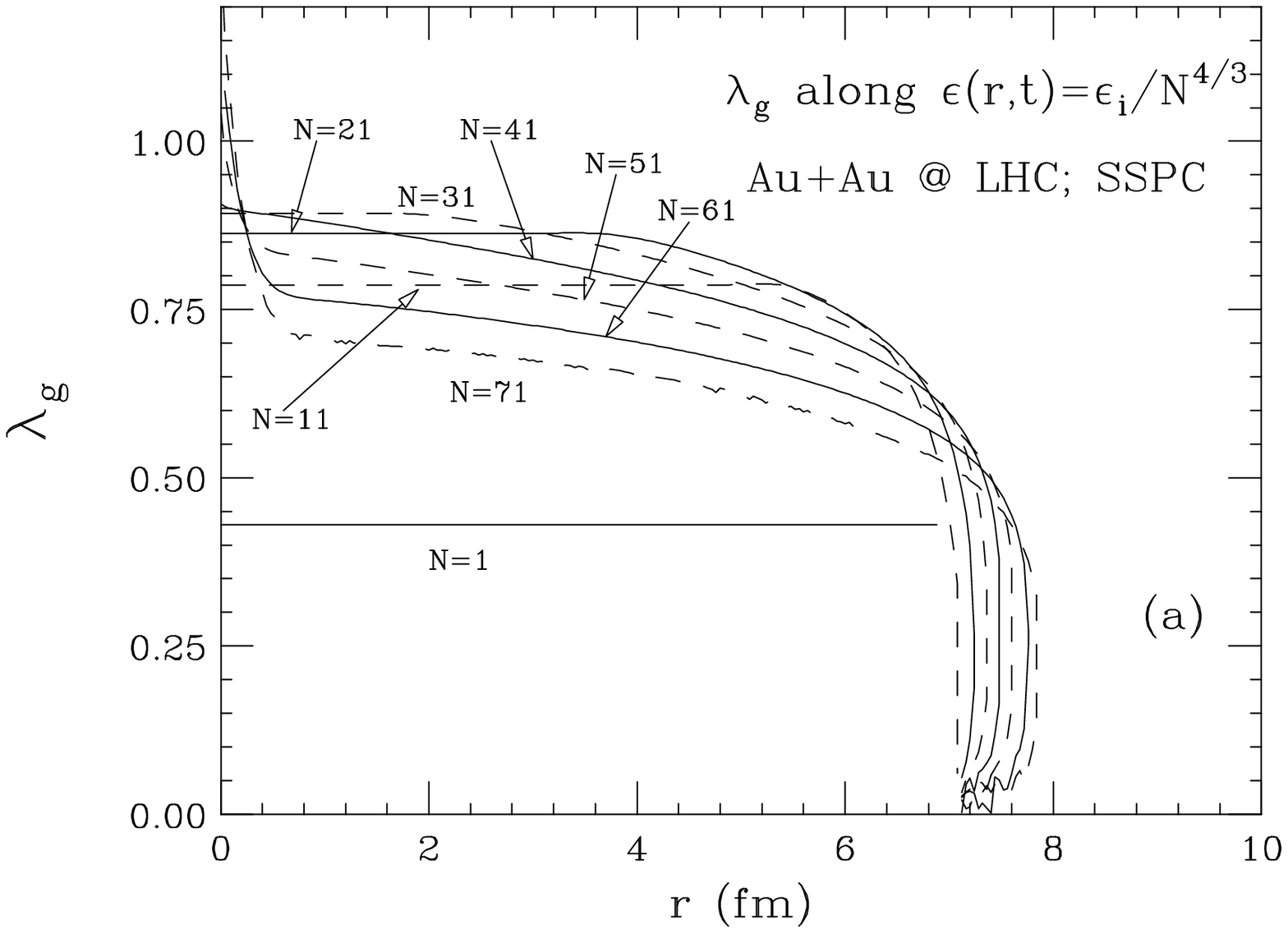}
\vskip 0.10in
\epsfxsize=3.25in
\epsfbox{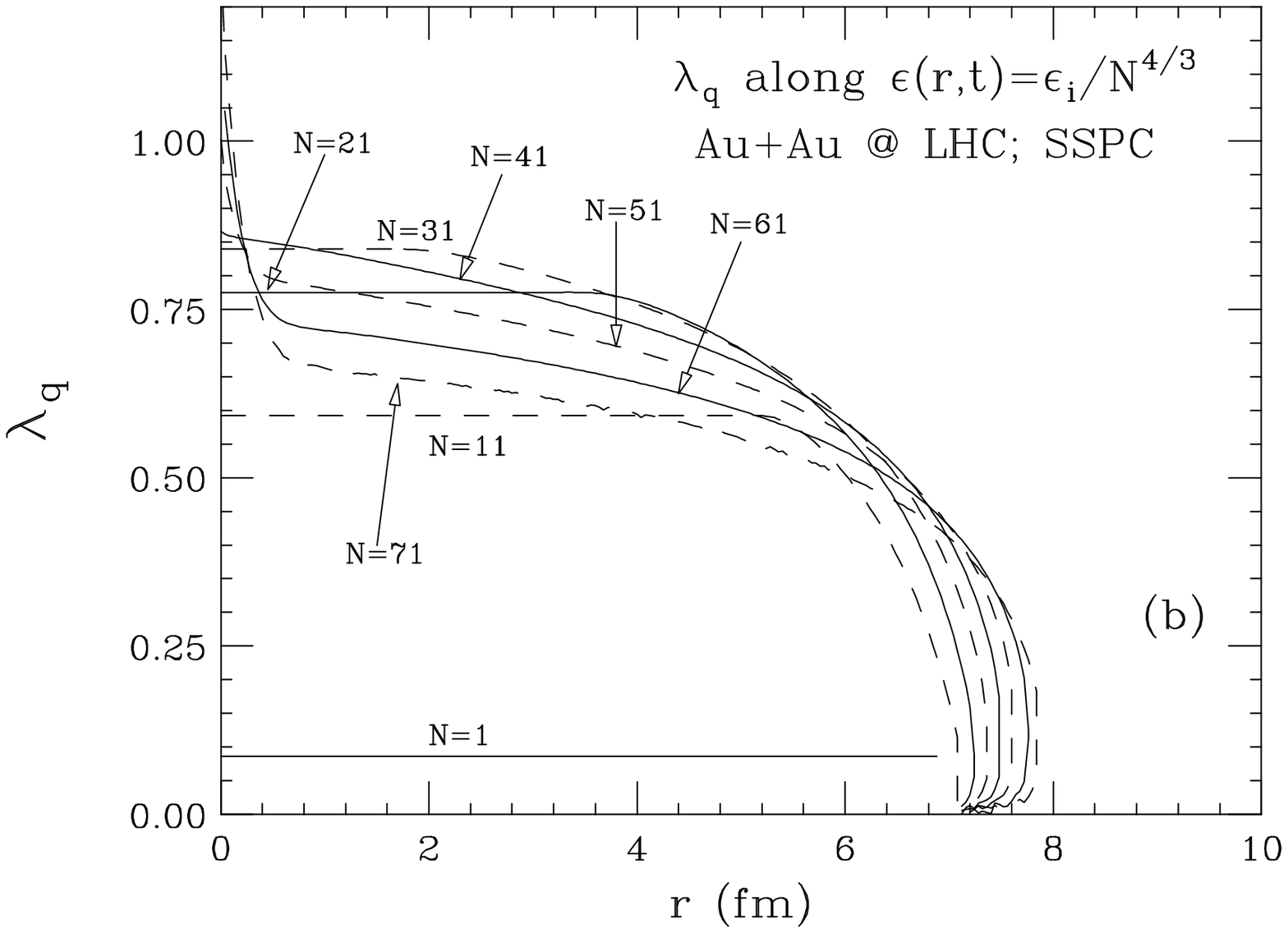}
\vskip 0.10in
\caption{ Same as before, at LHC.  }
\end{figure}

\section{PHOTON AND DILEPTON PRODUCTION FROM CHEMICALLY EQUILIBRATING
PLASMA}

In this section we wish to outline the rate of production of real and 
virtual photon due to bremsstrahlung in an chemically equilibrating 
plasma from the following reactions:
\begin{equation}
a \ + \ b \rightarrow c \ + \ d \ + \ \gamma (\gamma^\star) \ . 
\label{reaction}
\end{equation}
One can write the matrix element for this processes as
\begin{equation}
{\cal M} \ = \ e {\cal M}_0 J^\mu \epsilon_\mu ,  \label{matrix}
\end{equation}
where ${\cal M}_0$ is the matrix element for $ab\rightarrow cd$ and $J^\mu$
is the real(virtual) photon current \cite{haglin,dipali}. Using kinetic theory
one can write down the rate of production of a photon per unit time per unit
volume at a temperature $T$ as
\begin{eqnarray}
q_0\frac {dN}{d^4x \ d^3q} &=& g_{ab}\frac{1}{2(2\pi)^3} \int \prod_{i=a,b} 
\left ( \frac {d^3p_i} {(2\pi)^3 \ 2E_i} n_i(E_i,\lambda_i)\right ) 
\nonumber  \\
&\times& \ \int \prod_{j=c,d} \left ( \frac {d^3p_j} {(2\pi)^3 \ 2E_j} 
\left (1\pm n_j(E_j,\lambda_j)\right ) \right ) \nonumber  \\
&\times& (2\pi)^4  \delta (p_a  +  p_b  -  p_c  -  p_d - q) 
|{\cal M}|^2  , 
\label{rate}
\end{eqnarray}
where $g_{ab}=N_a N_b (2s_a+1)(2s_b+1)$ is the color and spin appropriate 
for the reactions in (\ref{reaction}). We have also considered 
the enhancement (suppression) in the exit channel corresponding to 
boson (fermion). The non-equilibrium distribution function is given in 
(\ref{befd}). The product of non-equilibrium distribution functions appearing 
in (\ref{rate}) can be decomposed in terms of equilibrium distributions  
as follows
\begin{eqnarray}
n_a n_b (1\pm n_c) (1\pm n_d) &=& (1\pm \lambda_c {\tilde n}_c)\big [ 
\lambda_a \lambda_b \lambda_d \left (1\pm {\tilde n}_d \right ) 
{\tilde n}_a {\tilde n}_b \nonumber \\
&+&  \lambda_a \lambda_b \left (1- \lambda_d \right ) 
{\tilde n}_a {\tilde n}_b \big ] \ . \label{decomp}
\end{eqnarray}

As discussed earlier we neglect the $q$ in the $\delta$-function
in the above (\ref{rate}). To compensate this one needs to insert a phase 
space 
correction factor $\Phi$, defined as the ratio of three-body phase space 
to two-body phase space. This can be obtained~\cite{dip,dipali,thesis} as
\begin{equation}
\Phi \left (s,s_2,m^2_a,m^2_b\right ) = \frac{s\xi^{1/ 2} \left (s_2, 
m^2_a, m^2_b \right )} {s_2 \xi^{1/ 2} \left (s, m^2_a, m^2_b\right )}  \
 \ , \label{correction} 
\end{equation}
where $\xi(x,y,z)=x^2 - 2x(y+z)+(y-z)^2$,  
$s_2 = s-2q_0\sqrt s$ for real photon and $s_2 = s+M^2 -2q_0 \sqrt s$ 
for dileptons in the case of equal mass scattering ($m_a=m_c$ and $m_b=m_d$). 

Taking into account the non-equilibrium nature of the plasma, the phase space 
correction, and also inserting $\int ds \ \delta ( s - (p_a+p_b)^2)$,
the dilepton transverse mass spectrum yield due to bremsstrahlung of
a virtual photon 
($ab \rightarrow cd \gamma^\star \rightarrow cd\,l^+l^-$)
in the quark phase is obtained from (\ref{rate}) as 
\begin{eqnarray}
\frac{dN}{dM^2d^2M_Tdy}&=&\int\,\tau \,d\tau\, r\, dr\, d\phi\,
d\eta \ \frac{T^6g_{ab}}{16\pi^4} \nonumber \\
&\times& \int_{z_{min}}^{\infty} \,dz 
\frac{\xi(z^2T^2,m_a^2,m_b^2)} {T^4} \ 
 \left (1\pm \lambda_c {\tilde n}_c\right ) \nonumber\\
&\times &
\Big [ 
\lambda_a \lambda_b \lambda_d \sum_{n=0} \left(\pm 
e^{\beta E_c}\right )^n \frac{{\cal K}_1\left(z(n+1)\right )}{n+1} 
\nonumber \\ 
&+&\lambda_a \lambda_b \left (1-\lambda_d\right)  {\cal K}_1(z) \Big ] 
\Phi(s,s_2,m_a^2,m_b^2) \nonumber \\
&\times&
\left (E\frac{d\sigma_{ab\rightarrow cd}^{l^+l^-}}{dM^2d^3q}\right ),
\label{dilepton}
\end{eqnarray}
where $z_{min}= (m_a+m_b+M)/T$, $z=\sqrt {s}/T$ and 
$E_c=\sqrt{p^2_c+m^2_a}$ with 
$|{\vec p}_c|= \xi^{1/2}(s,m^2_a,m^2_b)/(2\sqrt {s})$. The
 cross-section for the process 
$ab \rightarrow cd \,l^+l^-$
is given by
\begin{equation}
E\frac{d\sigma_{ab \rightarrow cd}^{l^+l^-}}{dM^2d^3q} =
\frac{d\sigma_{ab \rightarrow cd}^{l^+l^-}}{dM^2d^2M_Tdy} =
\frac{\alpha^2}{12 \pi^3 M^2} \,\frac{\widehat{\sigma}(s)}
{M_T^2\,\cosh^2 y}, \label{dixs}
\end{equation}
where $M_T$ is the transverse mass of the dilepton and $y$ is its
rapidity, so that, $q_0=M_T\,\cosh y$. The ${\widehat \sigma}(s)$ is
defined as
\begin{eqnarray}
{\widehat \sigma}(s) &=& \int^0_{-\xi(s,m^2_a,m^2_b)/s}\ dt \ 
\frac {d\sigma_{ab\rightarrow cd}}{dt} \ \nonumber \\
&\times& \left (q^2_0 |\epsilon \cdot J|^2_{ab\rightarrow cd} \right )
 \ . \label{txs}
\end{eqnarray}

The corresponding result for the transverse mass distribution
of dileptons from the equilibrating plasma, due to the annihilation process,
is given \cite{munshi} by
\begin{eqnarray}
\frac{dN_{\l^+\l^-}}{dM^2\,d^2M_T\,dy}=\frac{\alpha^2}{2\pi^3}
& &\lambda_{q}\lambda_{\bar{q}}
e_q^2 \int \tau\,d\tau\,rdr\,\nonumber\\
& &I_0\left(\frac{\gamma v_r p_T}{T}\right)
K_0\left(\frac{\gamma M_T}{T}\right). \ \ \label{trandis}
\end{eqnarray}
We would also like to point out that we consider the soft dilepton production
in the low invariant mass region ($M< m_{\rho}$). There are many sources of 
dilepton in this mass range. If quark-gluon plasma is formed, there will be
$q\bar q$ annihilation and $qq (\bar q)$ scattering with virtual 
bremsstrahlung. It has been shown in Ref.~\cite{alther} that in the case of
low mass pairs the perturbative $\alpha_s$-correction can be larger than
the $q\bar q$ annihilation spectra, which has been calculated in the Born
approximation. So, we consider this $q\bar q$ annihilation spectra as a lower 
limit and compare our result with it. 

The yield of soft photons produced from bremsstrahlung process,
$ab\rightarrow cd\gamma$ in an equilibrating plasma can also be obtained
from (\ref{rate}) as
\begin{eqnarray}
\frac{dN}{d^2q_Tdy}&=&\int\,\tau \,d\tau\, r\, dr\, d\phi\,
d\eta \ \frac{T^6g_{ab}}{16\pi^4} \nonumber \\
&\times& \int_{z_{min}}^{\infty} \,dz 
\frac{\xi(z^2T^2,m_a^2,m_b^2)} {T^4} \ 
 \left (1\pm \lambda_c {\tilde n}_c\right ) \nonumber\\
&\times &
\Big [ 
\lambda_a \lambda_b \lambda_d \sum_{n=0} \left(\pm 
e^{\beta E_c}\right )^n \frac{{\cal K}_1\left(z(n+1)\right )}{n+1} 
\nonumber \\ 
&+&\lambda_a \lambda_b \left (1-\lambda_d\right)  {\cal K}_1(z) \Big ] 
\Phi(s,s_2,m_a^2,m_b^2) \nonumber \\
& \times& 
\left (q_0\frac{d\sigma_{ab\rightarrow cd}^{\gamma}}{d^3q}\right ),
\label{photon}
\end{eqnarray}
where $z_{min}= (m_a+m_b)/T$. The cross-section for the emission 
of a soft real photon is
\begin{equation}
q_0\frac{d\sigma_{ab\rightarrow cd}^{\gamma}}{d^3q} = \frac{\alpha}{4\pi^2}\,
\frac{{\widehat \sigma}(s)}{q^2_0} \ , \label{phxs}
\end{equation}
where ${\widehat \sigma}(s)$ is defined as before (\ref{txs}) with $J^\mu$
replaced by the real photon current~\cite{dipali}. 

The photon production due to annihilation ($q\bar q \rightarrow g\gamma$) 
and Compton ($q(\bar q)g\rightarrow q(\bar q) \gamma$) processes in QGP
have already been studied in fair detail by number of
authors~\cite{kapus,traxler,baier,stric}, in which thermal masses have 
been used to shield the singularities in the scattering cross-section. 
It has been shown~\cite{kapus} that this is equivalent to using the 
resummation method of Braaten and Pisarski~\cite{pisarski} to regulate
the divergences of the QCD rates for these processes. In the following
we shall use the result of Ref.~\cite{traxler} for a 
chemically equilibrating plasma for comparison:

\begin{eqnarray}
\frac{dN}{d^2q_Tdy}&=&\int\,\tau \,d\tau\, r\, dr\, d\phi\,
d\eta \, \frac{5 \alpha \alpha_s}{18\pi^4} T^2 e^{-E/T}\nonumber \\
&\times& \Bigl[\lambda^2_q \lambda_g \pi^2
\Big[\ln\left(\frac{4ET}{k^2_c}\right)-1.42 \Big] \nonumber\\
&+&2\lambda_q \lambda_g \left(1 - \lambda_q \right) \Big[1-2 \gamma +2
\ln\left(\frac{4ET}{k^2_c}\right) \Big] \nonumber\\
&+&2\lambda_q \lambda_q \left(1 - \lambda_g \right) \Bigl[-2-2 \gamma +2
\ln\left(\frac{4ET}{k^2_c}\right) \Bigr] \Bigr] \, \label{traxler}
\end{eqnarray}

In order to make a comparison of our calculation with that of 
using HTL~\cite{auren} at the two loops level, we would like 
to discuss the following points.
As discussed in Ref.~\cite{auren,auren1}, the HTL calculation leads also 
to a separation in two factors: the amplitude square of the scattering 
process without photon emission, and a factor which is nothing but the 
square of an electromagnetic current responsible for photon emission. Now
one can easily convince oneself that this
essentially amounts to the soft photon approximation we have used in our
calculation. This can also be seen~\cite{auren} quantatively when one 
compares the bremsstrahlung production of a soft virtual photon in the HTL 
calculation with that of the semi-classical approximation by Cleymans 
et al~\cite{cley}. The semi-classical result agrees with the HTL result
in its functional dependence but over-estimates the rate of production by
$50\%$. This difference appears to be due to several simplifying assumptions
made in Ref.~\cite{cley}: (1) the photon momentum compared to the momenta of 
the constituents in the plasma has been neglected, but no phase space 
correction factor has been included. It has been discussed in 
Ref.~\cite{dip,dipali} that the result grossly over-estimates
without the phase space correction; (2) the approximate electromagnetic 
current has been used; (3) also simply a factor unity in the exit channel 
has been assigned instead of considering the appropriate statistical factors. 
With proper inclusion, like in the present calculation, of these factors should 
lead to an agreement with the thermal field theory result. However, the 
contribution obtained in HTL using two loops is rather large compared to
that in one loop~\cite{baart,wong}, though both are of the same order in 
the coupling constant.

The real photon production in a QGP within the two loops using 
HTL~\cite{auren} are also of the same order in the coupling constant as 
that of one-loop
but there is quantative difference corresponding to different
physical processes. The bremsstrahlung process
in two loops gives a contribution which 
is similar in magnitude to the Compton and
annihilation contributions evaluated up to the order of one loop~\cite{kapus}.
In two loops there is also an entirely new mechanism for the production of
hard photon through the annihilation of an off-mass shell quark and antiquark,
where the off-mass shell quark is a product of the scattering with another 
quark
or gluon. The contribution from this process completely dominates the emission
of hard photons as energy increases, and it is higher by more than one 
order of magnitude
for large energies~\cite{frank,dksph}. In Ref.~\cite{dksph} it has also been
shown that at the energy regime in which we are interested 
the bremsstrahlung from soft approximation
is little higher than that from two loops, but the contribution from soft
approximation falls off rapidly with increasing energy.

\section{RESULTS AND DISCUSSION}

The parton fugacities for an expanding plasma are 
displayed in Figs.(1-2) in Sec.II with initial conditions obtained from 
SSPC model~\cite{sspc}. One can see that
our results for the nonequilibrium soft photon~(\ref{photon}) and low
mass dileptons~(\ref{dilepton}) due to bremsstrahlung differ from that of
equilibrium rates. We would like to point out here that we cannot make 
a comparison of nonequilibrium results with that of equilibrium one as
the present study is intended for RHIC and LHC energies. The SSPC 
model~\cite{sspc} predicts that the QGP likely to be formed in this 
energies will be far away from chemical equilibrium. In view of this, 
comparing the nonequilibrium results with that of equilibrium one will 
also be far away from reality. However, a close inspection suggests 
that our results are, indeed, quite different.

In Fig.~3, we give our results for the transverse mass distribution of
low mass dileptons due to the bremsstrahlung process (solid curves) for 
$M =$ 0.1 GeV, 0.3 GeV and 0.5 GeV, respectively, at both RHIC and LHC 
energies for an equilibrating plasma. We have also given the contribution 
from quark-antiquark annihilation process (dashed curve) for a comparison. 
We see that the annihilation contribution is essentially the same 
for $M =$ 0.1 GeV, 0.3 GeV, and 0.5 GeV, thus representing a single 
curve in this figure. This is due to the fact that at $M$ below the 
$\rho$ mass 
the transverse mass distribution (see Eq.(\ref{trandis})) has a very
weak dependence on  $M$. But on the other hand it has a substantive 
dependence on the transverse mass $M_T$ and thus shows $M_T$ 
scaling \cite{asa,dks}. This scaling is broken when we take into account 
the bremsstrahlung contribution from the QGP sector. We now see that in the 
region $M\le 0.3$ GeV, the bremsstrahlung contribution 
out-shine the annihilation contribution at all $M_T$, while at $M \ge 0.3 $ 
GeV, the distribution is dominated by the annihilation contribution 
with increasing available energy and invariant mass. This is because of the 
fact that quark fugacities are seen to lag behind the 
gluon fugacities at all time and all radial distances. Since our entire 
study is based on the description where the plasma as initially produced 
is gluon rich, the yield would be dominated by the quark-gluon 
bremsstrahlung. In this context, we see that the relative importance 
of bremsstrahlung contribution increases quite considerably for a chemically
non-equilibrium plasma.

\begin{figure}
\epsfxsize=3.25in
\epsfbox{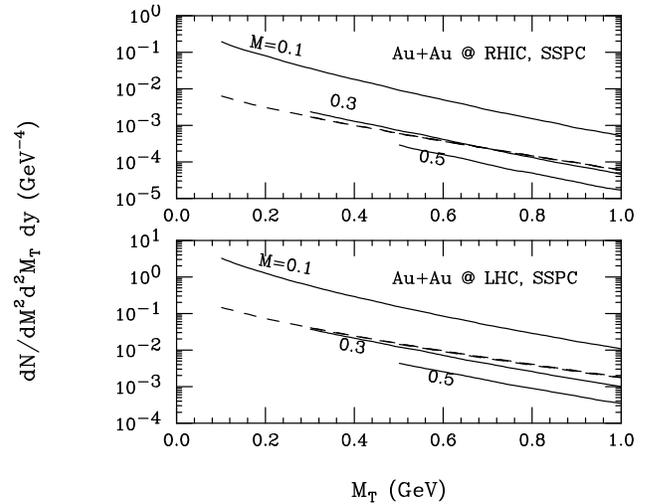}
\vskip 0.2 in
\caption{The transverse mass distribution of low mass dileptons 
calculated with quark driven bremsstrahlung (solid curve) and quark-
anti-quark annihilation (dashed curve; $M_T \geq M$) for $M =$ 0.1 GeV, 
0.3 GeV, and 0.5 GeV respectively at both RHIC and LHC energies.}
\end{figure}

We also plot the invariant mass spectrum of dileptons produced 
from bremsstrahlung (solid curve) and annihilation (dashed curve)
processes in fig.~4. At $M \leq 0.3 $ GeV, the bremsstrahlung process 
is seen to play the dominant role at both RHIC and LHC energies. 
We find that the low mass dilepton yield increases by a factor of 
20 as we go from RHIC to LHC energies. 

\begin{figure}
\epsfxsize=3.25in
\epsfbox{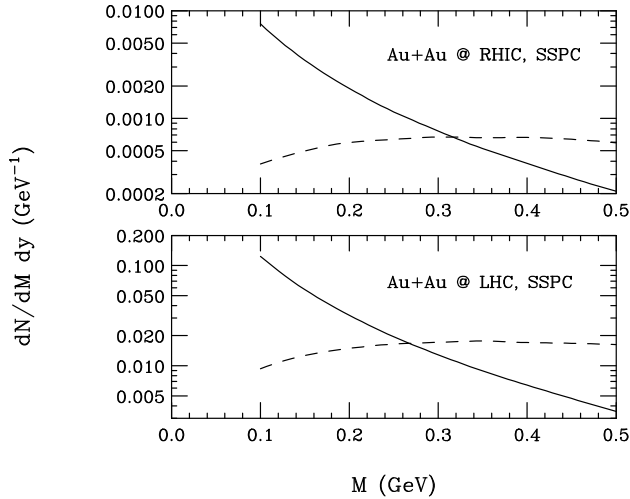}
\vskip 0.2 in
\caption{The invariant mass distribution of low mass dileptons at 
both RHIC and LHC energies. The solid line represents the contribution 
from bremsstrahlung process in quark phase and dashed line represents 
the annihilation contribution from quark phase.}
\end{figure}

Figure 5 shows the transverse momentum distribution of soft photons at 
RHIC and LHC energies for an equilibrating plasma. We find that the 
contribution of the bremsstrahlung process (solid curve) dominates over 
that of Compton and annihilation process (dashed curve) up to $p_{T} \leq 
0.5$ GeV at RHIC energy and at LHC energy, this dominance exists for $p_{T} 
\leq 0.6 $ GeV, after which they fall rapidly. We envisage 
an increase by a factor of almost 10 in the soft-photon yield as we move 
from RHIC to LHC energies. For the low $p_T$ regime, the photon rate has 
a substantial contribution from bremsstrahlung process
whereas in the high $p_T$ regime it is due to the hard 
processes (Compton and annihilation). As discussed already it will really be 
very interesting to study
the production of single photons from QGP using the two loops
results of Aurenche et al~\cite{auren} along with the one-loop contribution
in a chemically equilibrating plasma, since it has been 
suggested~\cite{sspc} that a chemically equilibrated plasma is likely
to be produced in RHIC and LHC energies with a fairly large temperature.
This indicates that QGP likely to be produced at those energies will have
a larger life time, and most likely the photons from the quark matter with 
the new rates~\cite{auren} may out-shine the photons from hadronic matter.

\begin{figure}
\epsfxsize=3.25in
\epsfbox{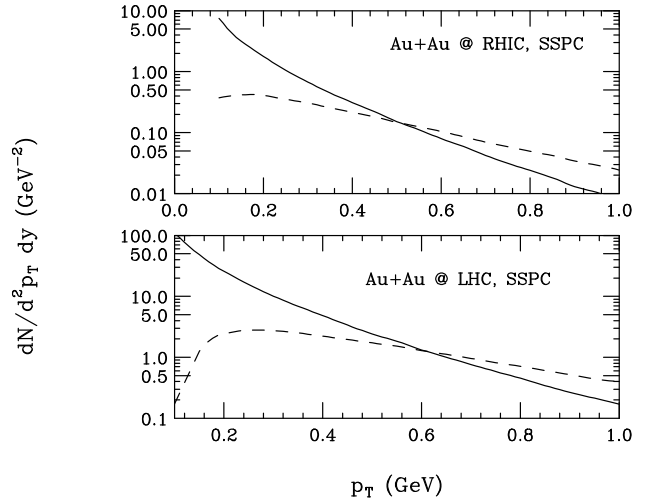}
\vskip 0.2 in
\caption{The transverse momentum distribution of soft photons at both
RHIC and LHC energies.}
\end{figure}

\section{SUMMARY}
In summary, we have studied  low mass dileptons and soft photons 
production due to bremsstrahlung process at both RHIC and LHC 
energies from an equilibrating and transversely expanding quark-
gluon plasma. We have chosen initial conditions from SSPC 
model in which the system is assumed to be in kinetic 
equilibrium at the proper time $\tau_{i} = 0.25 $ fm/c but far away
from chemical equilibrium \cite{geiger}. In the present study, we 
restrict ourselves to photons and dileptons having energies larger 
than $100$ MeV so that the Landau Pomeranchuk suppression \cite{landau} 
may not be substantial there. For dileptons, the bremsstrahlung 
contribution is seen to dominate up to $M \leq 0.3$ GeV and for 
photons this dominance exists for $p_{T} \leq 0.5$ GeV. We also find an 
increase by a factor of 15-20 in the low mass dilepton and 
soft photon yield as we go from RHIC to LHC energies where the 
life-time of plasma phase is expected to be large.

\vskip 0.3in

\centerline {ACKNOWLEDGMENT}

\vskip 0.3in

The authors are grateful to Dinesh Kumar Srivastava for
valuable suggestions and extensive discussions during the course
of this work. MGM is also thankful to the Alexander von Humboldt foundation,
Bonn, Germany,
for financial support.

\end{document}